# QCD Thermodynamics at $N_t = 8$ and 12

C. Bernard [a], T. Blum [b], T. DeGrand [c], C. DeTar [d], S. Gottlieb [e], U.M. Heller [f], L. Kärkkäinen [b], A. Kennedy [f], S. Kim [g], J. Kogut [h], R. Renken [i], K. Rummukainen [e], D.K. Sinclair [g], R. Sugar [j], Doug Toussaint [b] and K. Wang [k] *

[a]Department of Physics, Washington University, St. Louis, MO 63130, USA

[b]Department of Physics, University of Arizona, Tucson, AZ 85721, USA

[c]Physics Department, University of Colorado, Boulder, CO 80309, USA

[d]Physics Department, University of Utah, Salt Lake City, UT 84112, USA

[e]Department of Physics, Indiana University, Bloomington, IN 47405, USA

[f]SCRI, The Florida State University, Tallahassee, FL 32306-4052, USA

[g]HEP Division, Argonne National Laboratory, 9700 S. Cass Ave., Argonne, IL 60439, USA

[h]Department of Physics, University of Illinois, 1110 W. Green St., Urbana, IL 61801, USA

[i]Department of Physics, University of Central Florida, Orlando, FL 32816, USA

[j]Department of Physics, University of California, Santa Barbara, CA 93106, USA

[k]School of Physics, University of New South Wales, PO Box 1, Kensington, NSW 2203, Australia

We present results from studies of high temperature QCD with two flavors of Kogut-Susskind quarks on $16^3 \times 8$ lattices at a quark mass of $am_q = 0.00625$ and on $24^3 \times 12$ lattices at quark masses $am_q = 0.008$ and 0.016. The value of the crossover temperature is consistent with that obtained on coarser lattices and/or at larger quark masses. Results are presented for the chiral order parameter and for the baryon number susceptibility.

In this note we report on two studies of high temperature QCD with Kogut-Susskind quarks. The HTMCGC collaboration has recently completed a study on $16^3 \times 8$ lattices at a quark mass of $am_q = 0.00625$ where $a$ is the lattice spacing[1]. This project is a follow-on to earlier work on the same sized lattices at $am_q = 0.0125$[2]. The MILC Collaboration is in the process of carrying out calculations on $24^3 \times 12$ lattices with quark masses $am_q = 0.008$ and 0.016. The object of these projects is to push the study of high temperature QCD towards the continuum limit and physical quark masses in order to determine the nature of the transition between ordinary hadronic matter and the quark–gluon plasma, and to make more reliable calculations of the transition temperature and other thermodynamic quantities.

In the $N_t = 8$ project we have performed simulations for gauge couplings in the range $5.45 \leq 6/g^2 \leq 6.00$. A total of 1,000 trajectories (molecular dynamics time units) were generated for each value of the coupling studied except $6/g^2 = 5.475$ for which 1,450 were obtained. In Figure 1 we show the average value of the chiral order parameter, $\langle \bar\psi\psi \rangle$, and the baryon number susceptibility, $\chi$, obtained from these runs as a function of $6/g^2$. As expected, $\chi$ is small in the low temperature regime since a baryon must be created or destroyed to change the baryon number, while it is large in the high temperature regime since there

---

*presented by R. Sugar

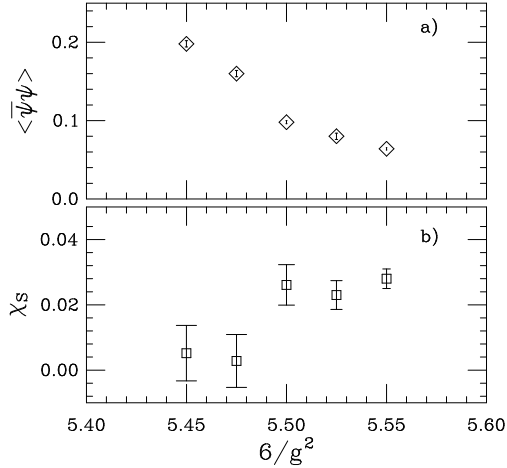

Figure 1: a) The chiral order parameter and b) the baryon number susceptibility as a function of $6/g^2$ on $16^3 \times 8$ lattices at $am_q = 0.00625$.

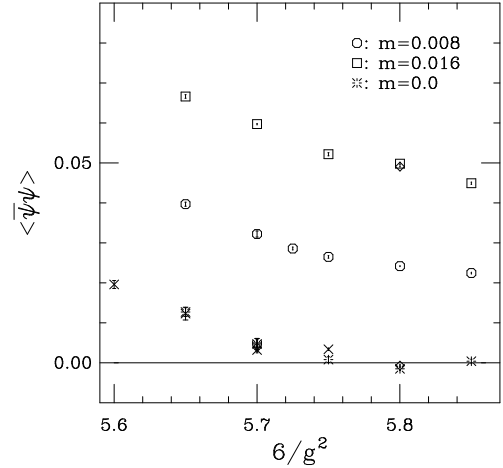

Figure 2: The chiral order parameter, $\langle\bar\psi\psi\rangle$ as a function of $6/g^2$ on $24^3 \times 12$ lattices.

it is only necessary to create or destroy quarks to change the baryon number. Just as on coarser lattices[3], the baryon number susceptibility provides an excellent signal for the crossover. We estimate the crossover value of the coupling at this quark mass to be $6/g^2 = 5.487(13)$

For $N_t = 12$ we are studying gauge couplings in the range $5.65 \leq 6/g^2 \leq 5.85$ for quark masses $am_q = 0.008$ and $0.016$. The autocorrelation times are large, and we plan to obtain approximately 2,000 trajectories at each coupling by the time this project is completed. At present we have over 1,000 trajectories at all couplings at which we are running except $6/g^2 = 5.85$, $am_q = 0.008$. However, the results presented below must be treated as preliminary.

In Figure 2 we plot the chiral order parameter, $\langle\bar\psi\psi\rangle$, as a function of $6/g^2$ for both of the masses which we are studying, $am_q = 0.008$ and $0.016$. We also show linear extrapolations to zero quark mass obtained from pairs of points at $am_q = 0.008$ and $0.016$. For some couplings extrapolations are shown from more than one pair of points. The bursts are extrapolations from two points with the same coupling, but different masses, while the fancy crosses are extrapolations from points with different couplings. Using results for the chiral order parameter and for the Polyakov loops, we estimate that for $am_q = 0.008$ the crossover occurs at $6/g^2 = 5.71(2)$. It is premature to estimate the position of the crossover for $am_q = 0.016$.

In Figure 3 we plot $T_c/m_\rho$ as a function of $(m_\pi/m_\rho)^2$. Here, $T_c$ is the temperature at the pseudocritical point determined in simulations run at finite quark mass:

$$T_c/m_\rho = 1/(N_t a m_\rho). \qquad (1)$$

In order to produce this figure we have made fits of $m_\pi$ and $m_\rho$ as a function of gauge coupling and quark mass, from spectrum calculations with two flavors of Kogut-Susskind quarks. These fits allow one to extrapolate $am_\pi$ and $am_\rho$ to the quark mass and pseudocritical value of the gauge coupling of the thermodynamics simulations. The non-Goldstone pion is used in this plot to point out the large flavor symmetry breaking. The line on the left hand side of the plot marks the physical value of $(m_\pi/m_\rho)^2$. Results with Wilson quarks are included for comparison. The curved line running through the $N_t = 8$ Wilson point shows the path it might fall on while staying within error bars. The $N_t = 12$ point is for

$am_q = 0.008$. It is quite preliminary, and it would be premature to attribute any significance to the fact that it is slightly larger than values of $T_c/m_\rho$ at larger lattice spacings. Rather, we consider it quite striking that there is such good agreement between the Kogut-Susskind results form $N_t = 4$ to $N_t = 12$, and between them and the Wilson results for $N_t = 6$ and 8.

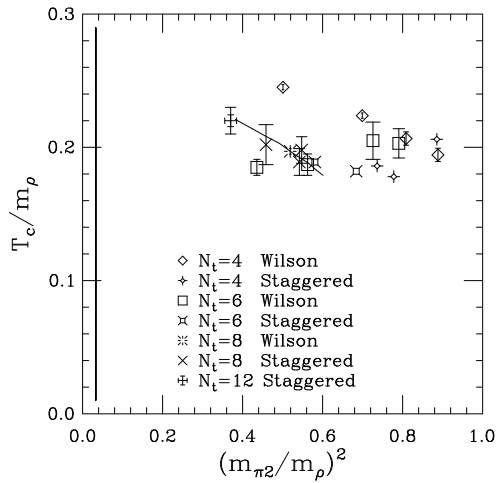

Figure 3: $T_c/m_\rho$ as a function of $(m_\pi/m_\rho)^2$. The non-Goldstone pion is used for Kogut-Susskind quarks.

The mass spectrum fits can also be used to calculate the baryon number susceptibility in physical units. This quantity enters into the QCD equations of state. It also plays an important role in models of nucleosynthesis, hadronization and heavy ion collisions. In Figure 4 we show the present world's data. The insensitivity of the results to lattice spacing is encouraging.

The scaling properties of thermodynamic quantities are important for determining the nature of the transition to the quark-gluon plasma. The results presented here are consistent with the expectation that for two flavors of quarks, there is a second order phase transition for zero quark mass, and a rapid crossover for finite quark masses[4]. This topic is discussed in some detail in DeTar's review talk[5].

The HTMCGC Collaboration's calculations at $N_t = 8$ were carried out on the Connection Machine CM2 at the Pittsburgh Supercomputer Center. The MILC Collaboration's $N_t = 12$ calculations are being performed on the Connection Machine CM5 at the National Center for Supercomputer Applications, Intel Paragons at the San Diego Supercomputer Center and Sandia National Laboratory, and Cray T3D's at the Pittsburgh Supercomputer Center and the National Energy Research Supercomputer Center. The work was supported by the NSF and DOE.

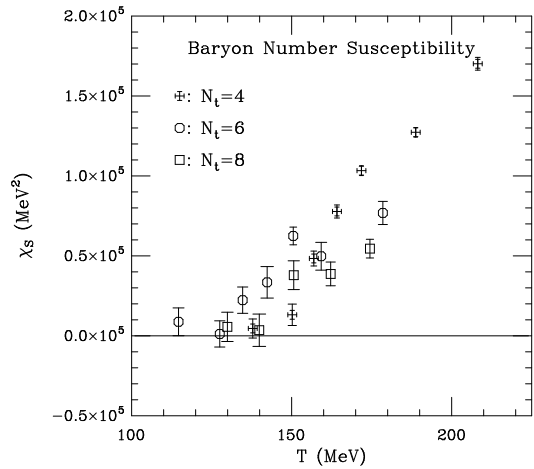

Figure 4: The baryon number susceptibility in physical units as a function of temperature.

## REFERENCES

1. The HTMCGC Collaboration, S. Gottlieb et al, Nuclear Physics B (Proc Suppl), **30** (1993) 315.
2. The HTMCGC Collaboration, S. Gottlieb et al, Phys. Rev. D, **47** (1993) 3619.
3. S. Gottlieb, W. Liu, D. Toussaint, R.L. Sugar and R.L. Renken), Phys. Rev. Letters, **59** (1987) 2247; Phys. Rev. D **38** (1988) 2888.
4. R. Pisarski and F. Wilczek, Phys. Rev. D **29** (1984) 338; F. Karsch, Phys. Rev. D **49** (1994) 338.
5. C. DeTar, this volume.